\def\ha{H$\alpha$}
\def\nt{[N\,{\sc ii}]}

\documentclass[aaspp4,11pt,preprint,apjfonts]{emulateapj}

\shorttitle{The Massive Hubble Sequence beyond $z=2$}
\shortauthors{Kriek et al.}

\begin{document}
  
\title{The Massive Hubble Sequence beyond $\lowercase{z}=2$ and the Connection between Large Star-Forming and Compact Quiescent Galaxies\altaffilmark{1}}

\author{Mariska Kriek\altaffilmark{2}, Pieter G. van
  Dokkum\altaffilmark{3}, Marijn Franx\altaffilmark{4}, Garth D.
  Illingworth\altaffilmark{5} \& Daniel K. Magee\altaffilmark{5}}

\altaffiltext{1}{Based on observations made with the NASA/ESA {\it Hubble Space Telescope (HST)}, obtained at the Space Telescope Science Institute, which is operated by AURA, Inc., under NASA contract NAS 5--26555.}

\altaffiltext{2}{Department of Astrophysical Sciences, Princeton University, 
 Princeton, NJ 08544, USA}

\altaffiltext{3}{Department of Astronomy, Yale University, New Haven, 
  CT 06520, USA}

\altaffiltext{4}{Leiden Observatory, Leiden University, NL-2300 RA Leiden, 
  Netherlands}

\altaffiltext{5}{UCO/Lick Observatory, University of California, Santa
  Cruz, CA 95064, USA}

\begin{abstract} 
  We present {\it Hubble Space Telescope} NIC2 morphologies of a
  spectroscopic sample of massive galaxies at $z\sim2.3$, by extending
  our sample of 9 compact quiescent galaxies ($r_e\sim0.9$\,kpc)
  with 10 massive emission-line galaxies. The emission-line galaxies
  are classified by the nature of their ionized emission; there are
  six star-forming galaxies and four galaxies hosting an active
  galactic nucleus (AGN). The star-forming galaxies are the largest
  among the emission-line galaxies, with a median size of
  $r_e=2.8$\,kpc. The three galaxies with the highest star formation
  rates ($\gtrsim 100\,M_{\odot}\rm\,yr^{-1}$) have irregular and
  clumpy morphologies. The AGN host galaxies are more similar to the
  compact quiescent galaxies in terms of their structures ($r_e \sim
  1.1\,$kpc) and spectral energy distributions. The total sample
  clearly separates into two classes in a color--mass diagram: the
  large star-forming galaxies that form the blue cloud, and the
  compact quiescent galaxies on the red sequence. However, it is
  unclear how or even if the two classes are evolutionary
  related. Three out of six massive star-forming galaxies have dense
  cores and thus may passively evolve into compact galaxies due to
  fading of outer star-forming regions. For these galaxies a reverse
  scenario, in which compact galaxies grow inside-out by star
  formation is also plausible. We do caution though that the sample is
  small. Nonetheless, it is evident that a Hubble sequence of massive
  galaxies with strongly correlated galaxy properties is already in
  place at $z>2$.
\end{abstract}


\section{INTRODUCTION}\label{sec:intro}

Theoretical models are slowly converging to a coherent picture on how
massive galaxies are building up their stellar mass over cosmic
time. The early phase is thought to be comparable to a dissipative
collapse model, and dominated by mergers of gas-rich sub-components
and in situ star formation \citep[e.g.,][]{na07}. This cold accretion
mode is efficient for low-mass halos, but is expected to cease once
the halo becomes too massive and too hot
\citep[e.g.,][]{bi07}. However, massive galaxies at high redshift may
still grow by cold clumpy streams, which penetrate the halo through
the filaments \citep[e.g.,][]{de08,de09}. In the second ``quiescent''
phase the galaxy is thought to primarily grow by accretion of smaller
systems, and will build up at larger
radii \citep[e.g.,][]{na07,na09,bo07,be09}.

This picture is primarily inspired by new observational
results. First, several studies find that $z>2$ star-forming galaxies
exhibit irregular and clumpy structures
\citep[e.g.,][]{el07,el09,ge08,fo09}. Second, massive quiescent
galaxies are found to increase in size by about a factor of $\sim5$
\citep[e.g.,][]{tr06,lo07,vd08,ci08,fr08} from $z\sim2$ to the
present, and this size growth is confirmed by the first dynamical studies
\citep{vd09b,ca09}. 

Nonetheless, there is no observational evidence for
an evolutionary relation between the irregular star-forming and compact
quiescent galaxies. Clumpy star-forming galaxies could form compact
bulges, as the clumps may migrate to the center and coalesce
\citep[][]{de09b,el08}. However, the high-redshift clumpy galaxies
detected so far are not massive enough to qualify as direct
progenitors of compact quiescent galaxies. In order to assess a
possible evolutionary sequence, detailed morphological studies of
complete massive galaxy samples are required.

In this Letter we present $HST$~NIC2 imaging of a spectroscopic sample
of massive galaxies at $z\sim2.3$. The galaxies without detected
emission lines were already studied in
\cite{vd08}. Here, we extend this study with the morphologies of
massive emission-line galaxies \citep{kr07}. This massive galaxy
sample enables us to study and relate the structures of star-forming
and quiescent galaxies, obtain a census of the ``Hubble Sequence'' at
$z\sim2.3$, and explore a possible evolutionary sequence. A
$\Lambda$CDM cosmology with $\Omega_{\rm m}=0.3$,
$\Omega_{\Lambda}=0.7$, and $H_{\rm0}=70$~km s$^{-1}$ Mpc$^{-1}$ and
AB magnitudes are assumed throughout.

\begin{figure} 
  \begin{center} \includegraphics[width=0.47\textwidth]{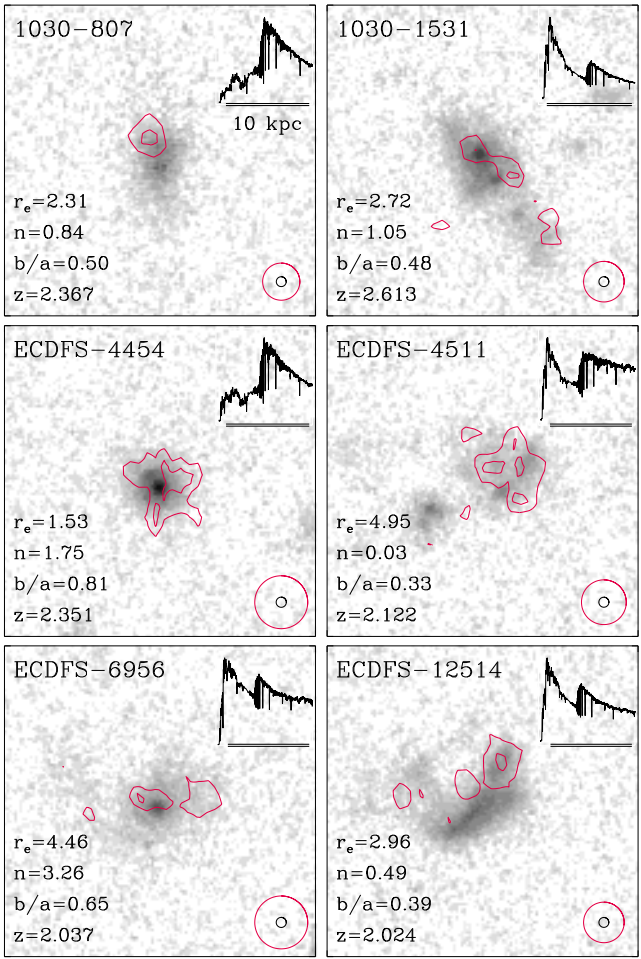}
  \caption{NIC2 $HST$ images of six massive star-forming galaxies at
  $z\sim2.3$. Overplotted in red are the $VLT$/SINFONI H$\alpha$+\nt\
  emission-line maps. The contours represent 60\% and 90\% of the
  maximum line emission in the galaxy. The NIC2 and SINFONI PSFs are
  presented by the black and red circles, respectively. The 10 kpc
  distance scale is indicated for each galaxy. The redshift and
  best-fit structural parameters ($r_e$ is circularized and in kpc)
  are printed as well. The average uncertainties on $r_e$ and $n$ are
  0.13 kpc and 0.05, respectively. These uncertainties only reflect
  photon noise. For each galaxy we show the best-fit stellar
  population model to the rest-frame optical spectrum and rest-frame
  UV photometry (in $f_{\lambda}$), in the observed wavelength range
  0.2-2.5 $\mu$m.\label{fig:sf}} \end{center}
\end{figure} 

\begin{figure} 
  \begin{center} \includegraphics[width=0.47\textwidth]{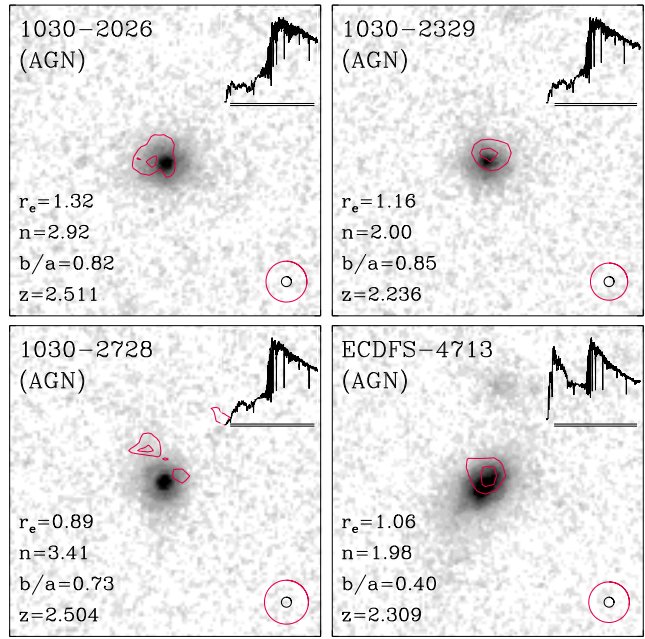}
  \caption{NIC2 $HST$ images of four massive AGN host galaxies at
  $z\sim2.3$. The average uncertainties on $r_e$ and $n$ are 0.03 kpc
  and 0.08, respectively. See the caption to Figure 1.\label{fig:agn}}
  \end{center}
\end{figure} 

\section{SAMPLE, OBSERVATIONS, REDUCTION AND FITTING}\label{sec:data}

The galaxies studied in this Letter are selected from our
spectroscopic sample of massive galaxies (${\rm log}\,M > 10.5$) at
$z\sim2.3$. For all galaxies we have spectroscopy with the $Gemini$
near-infrared (NIR) spectrograph \citep[GNIRS; ][]{el06} and deep
optical to NIR photometry as provided by the Multi-Wavelength Survey
by Yale-Chile \citep[MUSYC; ][]{ga06,qu07,ent09}. The full sample
consist of 28 massive galaxies at $2<z<3$, ranging from strong
starbursts, to AGN hosts, and quiescent systems. In terms of
rest-frame $U-V$ color this sample is representative for a
mass-limited sample \citep{kr08a}. More information on the galaxy
sample and properties of the individual galaxies can be found in
\cite{kr08a,kr08b} and \cite{mu09}.

We obtained $HST$ NIC2 imaging for 19 galaxies, the ones for which we
obtained GNIRS spectra early on. Consequently, the $HST$ NIC sample is
not representative for our full sample, and slightly biased to the
red. The $HST$ NIC2 morphologies of the nine galaxies without detected
emission lines \citep{kr06} were discussed in
\cite{vd08}. Here, we add $HST$ NIC2 imaging of the 10 massive
emission-line galaxies (Figures~\ref{fig:sf} and \ref{fig:agn}), with
$VLT$ SINFONI spectroscopy \citep{kr07}. They were observed from 2007
July -- 2008 October using the F160W filter. The total integration per
galaxy is three orbits, each split in two (dithered) exposures. The
data were reduced following the method by
\cite{bi06} and \cite{bo08}.

We measure the structural parameters for each galaxy by fitting a
single \cite{se68} radial surface brightness profile, using the
two-dimensional fitting code GALFIT \citep{pe02}, and allowing the
S\'ersic $n$-parameter to float. Thus, we assume all light is
stellar. We use Tiny Tim 6.3 \citep{kr95} to construct a synthetic
NIC2 point-spread function (PSF) for each galaxy
\citep[see][ for more details]{vd08}. The resulting fit parameters are
given in Figures~\ref{fig:sf} and \ref{fig:agn}.

\begin{figure*} 
  \begin{center} \includegraphics[width=0.95\textwidth]{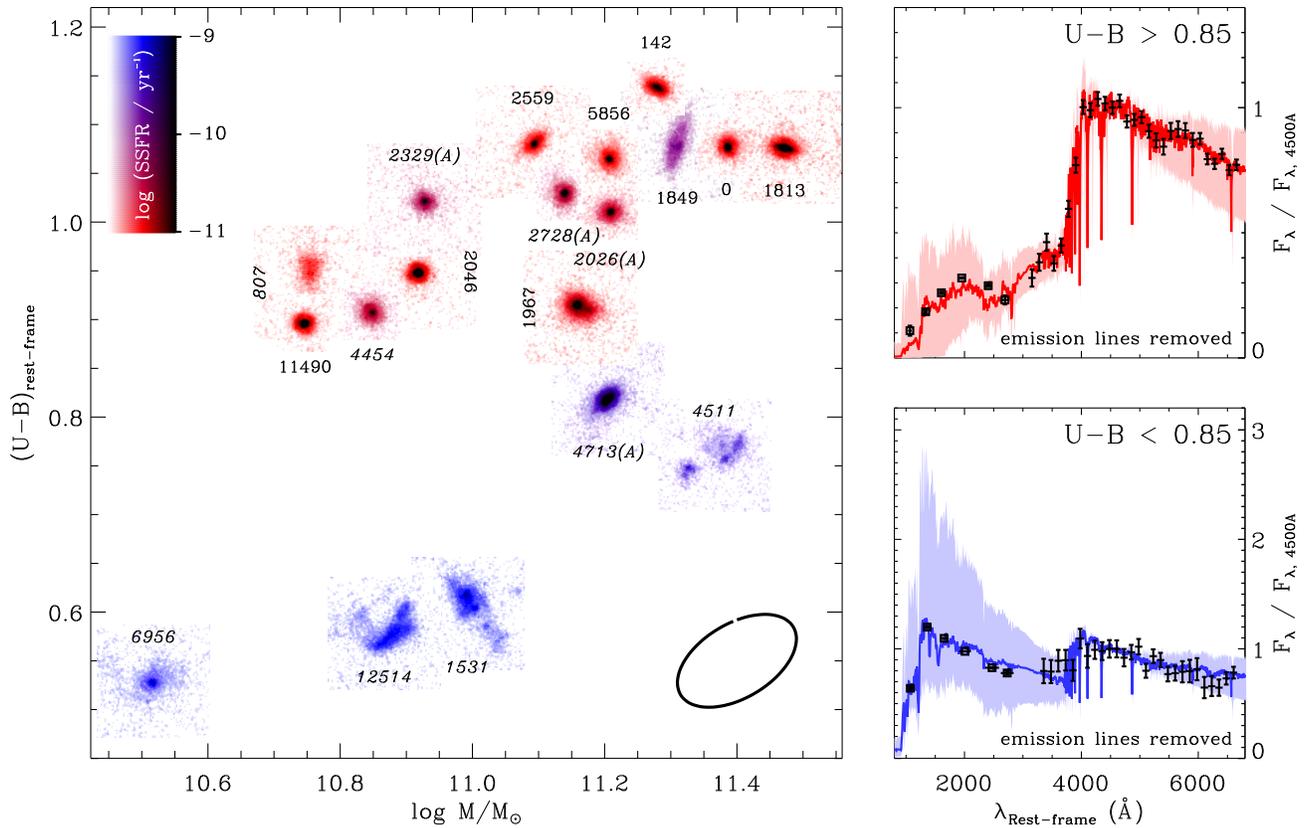} \caption{Left:
    rest-frame $U-B$ color vs. stellar mass for a massive galaxy
    sample at $z\sim2.3$ with rest-frame optical spectroscopy. We use
    the $HST$ NIC2 images as symbols. The color coding reflect the
    specific SFR of the galaxy. The emission-line galaxies can be
    recognized by their $italic$ ID numbers, and $A$ indicates the
    AGNs. The galaxies clearly separate into two classes: the large
    (irregular) star-forming galaxies in the blue cloud, and the
    compact, quiescent galaxies on the red sequence. We do caution
    that this sample is small and not complete. The ellipse represents
    the average 1\,$\sigma$ confidence interval.  Right: stacked SEDs,
    composed of the rest-frame UV photometry and rest-frame optical
    spectra of all blue (bottom panel) and red galaxies (top panel) at
    $2<z<3$ in our spectroscopic sample. We also show the stack and
    full range of best-fit stellar population synthesis (SPS)
    models. The SPS models do not have emission lines and thus they
    are correspondingly removed from the
    stacks. \label{fig:mass}}  \end{center} 
\end{figure*}

\section{THE HUBBLE SEQUENCE AT $z\sim2.3$}

Figures~\ref{fig:sf} and \ref{fig:agn} show that massive emission-line
galaxies at $2.0<z<2.7$ exhibit a wide range in morphologies. In
\cite{kr07} we use emission-line diagnostics to determine whether star
formation or active galactic nuclei (AGNs) dominate the line
emission. The six massive star-forming galaxies are the largest with a
median $r_e$ of 2.8~kpc, a median S\'ersic parameter of $n\sim1$, and
half (1030-1531, ECDFS-4511, and ECDFS-12514) show irregular and clumpy
morphologies. This fraction might be higher if we include the
line-emitting regions (\ha\ and \nt), as observed by $VLT$
SINFONI. The galaxies hosting AGNs have sizes ($r_e\sim1.1$\,kpc),
S\'ersic parameters ($n\sim2.5$), and spectral energy distributions
(SEDs) more similar to the compact, quiescent galaxies without
emission lines.

In Figure~\ref{fig:mass} we show the images of the full sample in a
color--mass diagram. The stellar masses, rest-frame colors, and star
formation rates (SFRs) are adopted from \cite{kr08a,kr08b,kr09}. The
SFRs are based on SED fits, not the H$\alpha$ luminosities. Stellar
masses and SFRs are corrected by 0.23 dex to convert from a
\cite{sa55} to a \cite{ch03} initial mass function. The color coding
reflects the specific SFR (SFR/$M_*$).

There is a striking correspondence between the location of galaxies in
the color--mass plane and morphology. The large (irregular) blue
galaxies make up the blue cloud. The emission-line galaxies 1030-807
and ECDFS-4454, with lower specific SFRs, are closer to the red
sequence and have sizes intermediate of the quiescent and star-forming
galaxies. Their line maps exhibit residual star-forming regions,
primarily in the outskirts. However, for ECDFS-4454 an AGN might
contribute to the line emission as well \citep{kr07}. Also HDFS1-1849
on the red sequence still has ongoing star formation according to the
SED fit, and is likely a dusty edge-on disk. The high dust content
might be the reason why we detected no emission lines for this
galaxy. Three compact AGN hosts join the quiescent compact galaxies on
the red sequence. The AGN host ECDFS-4713 is also compact, but has a
higher specific SFR. However, in contrast to the other AGNs which are
faint in the rest-frame UV, for this galaxy we cannot exclude that the
UV emission might be of nuclear origin. Altogether, this diagram
illustrates that structures and stellar population properties of
massive galaxies at $z\sim2.3$ are strongly correlated.

Our results confirm previous studies based on lower resolution $HST$
NIC3 or ground-based imaging of photometric galaxy samples
\citep[e.g.,][]{to07,zi07,fr08,wi09}. Moreover, due to the higher
spatial resolution, we better resolve the structures of the massive
star-forming galaxies. This work extends the results by
\cite{el07,el09} and \cite{fo09} to higher masses, suggesting that
star formation in irregular and clumpy galaxies may represent the
major star-forming mode beyond $z=2$.

The massive Hubble sequence at $z\sim2.3$ is quite different from that
in the local universe. First, quiescent galaxies are much more compact
than local early-type galaxies (ETGs) at similar mass. Second, the
galaxies with the highest SFRs ($\gtrsim 100\,M_{\odot} \rm\,yr^{-1}$)
in our sample (1030-1531, ECDFS-4511, and ECDFS-12514) have irregular
and clumpy structures, and thus do not resemble classical disk or
spiral galaxies. Massive irregular galaxies with such high specific
SFRs are very rare in the local universe. Star-forming galaxies
1030-807, HDFS1-1849, and ECDFS-6956 are structurally more similar to
local massive disk galaxies, but their SFRs are also lower ($\lesssim
25\,M_{\odot}\,\rm yr^{-1}$).

\begin{figure} 
  \begin{center} \includegraphics[width=0.48\textwidth]{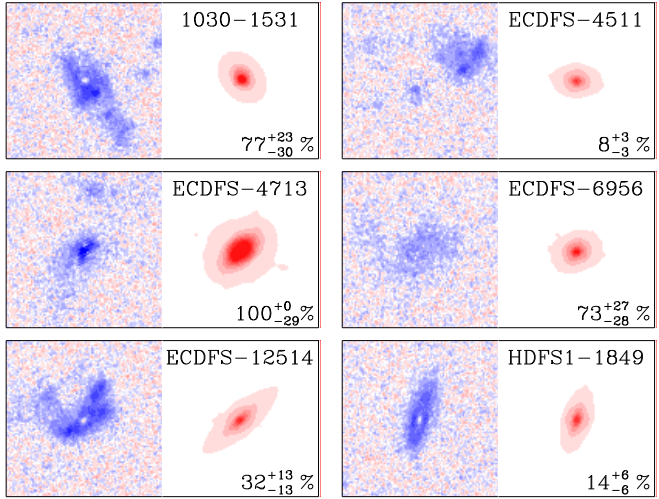}
    \caption{For each star-forming galaxy we show the maximum compact
    core ($right$) and the residual image after subtraction of this
    core ($left$). In the latter, blue and red represent positive and
    negative flux, respectively. The stellar mass fraction that could
    be in a dense core is given for each galaxy. \label{fig:sub}}
    \end{center}
\end{figure}

\section{DO BLUE GALAXIES HAVE DENSE CORES?}

An obvious question is whether the blue star-forming galaxies are
simply compact quiescent galaxies surrounded by active star-forming
regions or disk with much lower mass-to-light ratios ($M/L$). We
assess this by examining how much of the stellar mass could be in a
dense core for the six galaxies with the highest specific SFR in the
total sample. For the core we assume an $r_e$ of 0.9~kpc, an $n$ of 3
\citep{vd08}, and the median $M/L$ of the compact red-sequence
galaxies. We apply the appropriate PSF and fit the brightest clump or
core, leaving the axis ratio and the inclination as free
parameters. The maximum mass fraction is set to 100\% to ensure that
unrealistic values do not occur (since star-forming populations have
lower $M/L$). Figure~\ref{fig:sub} shows the compact cores and the
residual images. The uncertainties on the mass fractions are dominated
by variations in $M/L$ of the quiescent galaxies.

ECDFS-4713 , ECDFS-6956, and 1030-1531 can hide a major fraction of
their stellar mass in a compact core. Thus, aside from active outer
star-forming regions, they may be similar to the compact quiescent
galaxies. The remaining galaxies do not seem to have such a strong
light or mass concentration. However, this exercise is complicated by
the effects of dust. For example, HDFS1-1849 ($A_V$ = 1.6 mag) is
likely a dusty edge on disk, and so there could be a hidden compact
core.

In the above exercise we assume that the core has a $M/L$ similar to
the quiescent galaxies, and thus much higher than the $M/L$ of the
star-forming galaxies. If we were to assume a constant $M/L$ in the
star-forming galaxies, we would infer fractions of only 29\%, 76\%,
and 27\% for 1030-1531, ECDFS-4713 and ECDFS-6956, respectively. Thus,
it is crucial to study the stellar populations in the sub-components
in detail to examine whether they indeed exhibit strong $M/L$
variations.

\section{DISCUSSION} 

It is interesting to consider whether and how large star-forming and
compact quiescent galaxies at $z\sim2.3$ may be evolutionary
related. Below we discuss three different scenarios. The simplest
conceptually is the scenario in which massive blue galaxies stop
forming stars and move to the red sequence. In this scenario the
massive irregular star-forming galaxies may represent the early clumpy
star-forming phase, the red galaxies with residual star-forming
regions may be transitional objects, and the compact quiescent
systems represent the quenched phase in which galaxies live in hot
halos, and only grow by (minor) red mergers. In the previous section
we found that three massive star-forming galaxies might have 70\%--100\%
of the total stellar mass in a compact core, and thus simple fading of
outer star-forming regions could lead to a compact galaxy. However,
the three remaining star-forming galaxies need non-passive processes,
like (clump) coalescence in order to form a compact core. Irrespective
of the mechanism, this requires a large increase in the stellar
density of these systems.

Alternatively, compact quiescent galaxies may undergo periods of star
formation and move to the blue cloud during that time. In this case
galaxies may build up their outer parts by star formation. Several
studies indeed show that compact quiescent galaxies grow inside-out
(see Section 1), although the responsible process has yet to be
identified. For this scenario, the two red galaxies with residual star
forming regions, and the three star-forming galaxies with compact
cores may directly exemplify this inside-out growth. However, this
scenario fails for the star-forming galaxies lacking dense cores.

It is also possible that the two classes are not directly related, and
follow distinct evolutionary paths. Theoretical models find that ETGs
that form at later times are expected to be less compact
\citep[e.g.,][]{ks06}. Thus, the large blue galaxies at $z\sim2.3$ may
evolve into less compact ETGs.  The typical average surface density of
the star-forming galaxies presented here is $2 \times 10^{9}\,M_{\odot}
\rm \,kpc^{-2}$, about a factor of 10 lower than those of the quiescent
galaxies, but comparable to those of local ETGs. Thus the star-forming
galaxies at $z=2$ are likely progenitor of local ETGs, as was
previously found by \cite{fr08}. Moreover, if compact, quiescent
galaxies beyond $z=2$ evolve into the compact cores of local
elliptical galaxies, it may be plausible that the massive star-forming
galaxies evolve into the local S0's. The stars in these galaxies are
nearly as old as in local ellipticals, but their disk-like morphologies
separates them in a structural way.

It is also interesting to consider how AGNs tie in with a possible
evolutionary scenario. The fact that their hosts are more similar to
red sequence galaxies, although being slightly less quiescent and less
compact, may suggest that they are a transitional population. However,
the AGN activity may just as well follow the star-formation
activity, and the galaxy may be just temporarily ``flaring up''. In
this context it is interesting to note that by integrating
significantly deeper, 1255-0 hosted an AGN as well, albeit less
luminous \citep{kr09}. We do stress though that the AGN sample may
be incomplete, as our AGN identification is less appropriate for
star-forming galaxies \citep{kr07}.

Follow-up studies are needed to further discriminate between the
different scenarios. Clustering and dynamical studies will be
essential in linking the galaxy populations at different
redshifts. Moreover, we need to study the stellar populations of
different sub-components in massive $z>2$ galaxies. The
combination of $HST$ WFC3 and the NEWFIRM median-band survey
\citep[which provides accurate photometric redshifts and SEDs of large
samples of galaxies at $1<z<3$;][]{vd09a} will enable detailed studies
of larger samples of massive galaxies than presented here, to
link them to lower and higher redshift populations, and thus to follow
their star formation and assembly history over cosmic time.

\acknowledgements We thank Jeremiah Ostriker, Jenny Greene, and Jim Gunn
for useful discussions. Support for program HST-GO-11135.08 was
provided by NASA through a grant from the Space Telescope Science
Institute. G.D.I. and D.K.M. acknowledge support from NASA grant NAG5-7697.


\begin{thebibliography}{}
\bibitem[Bezanson et al.(2009)]{be09} Bezanson, R., van Dokkum, P., G.,
  Tal, T., Marchesini, D., Kriek, M., Franx, M., \& Coppi,
  P. 2009, \apj, 697, 1290
\bibitem[Birnboim et al.(2007)]{bi07} Birnboim, Y., \& Dekel, A., 
  Neistein, E. 2007, \mnras, 380, 339
\bibitem[Bournaud et al.(2007)]{bo07} Bournaud, F., Jog, C. J., 
  Combes, F. 2007, A\&A, 476, 1179
\bibitem[Bouwens \& Illingworth(2006)]{bi06} Bouwens, R. J., \&
  Illingworth, G. D. 2006, Nature, 433, 189
\bibitem[Bouwens et al.(2008)]{bo08} Bouwens, R. J., Illingworth, G. D., 
  Franx, M., \& Ford, H. 2008, \apj, 690, 1764
\bibitem[Cappellari et al.(2009)]{ca09} Cappellari, M. et al. 2009, 
  \apjl, submitted (arXiv:0906.3648)
\bibitem[Cenarro \& Trujillo(2009)]{tr09} Cenarro, A. J., \& Trujillo, I.
  2009, \apj, 696, L43
\bibitem[Chabrier(2003)]{ch03} Chabrier, G. 2003, \pasp, 115, 763
\bibitem[Cimatti et al.(2008)]{ci08} Cimatti, A., et al. 2008, A\&A,
  482, 21
\bibitem[Dekel \& Birnboim(2008)]{de08} Dekel, A., \& Birnboim,
  Y. 2008, \mnras, 383, 119
\bibitem[Dekel et al.(2009a)]{de09} Dekel, A., et al. 2009, Nature,
  457, 451
\bibitem[Dekel et al.(2009b)]{de09b} Dekel, A., Sari, R., Ceverino, D., 
  2009b, \apj, submitted (arXiv:0901.2458)
\bibitem[Elias et al.(2006)]{el06} Elias, J. H., et al. 2006, SPIE
  6269, 139
\bibitem[Elmegreen et al.(2007)]{el07} Elmegreen, D. M., Elmegreen,
   B. G., Ravindranath, S., \& Coe, D. A., 2007, \apj, 658, 763
\bibitem[Elmegreen et al.(2009)]{el09} Elmegreen, D. M., Elmegreen, 
  B. G., Marcus, M. T., Shahinyan, D., Yau, A., \& Peterson,
  M. 2009,  \apj, 701, 306
\bibitem[Elmegreen et al.(2008)]{el08} Elmegreen, B. G., Bournaud, 
  F., \& Elmegreen, D. M. 2008, \apj, 688, 67
\bibitem[F\"orster Schreiber et al.(2009)]{fo09} F\"orster Schreiber, 
  N. M. 2009, \apj, submitted (arXiv:0903.1872)
\bibitem[Franx et al.(2008)]{fr08} Franx, M., van Dokkum, P. G., 
  F\"orster Schreiber, N. M., Wuyts, S., Labb\'e, I., \& Toft, S. 
  2008, \apj, 688, 770
\bibitem[Gawiser et al.(2006)]{ga06} Gawiser, E., et al. 2006, \apjs,
  162, 1
\bibitem[Genzel et al.(2008)]{ge08} Genzel, R., et al. 2008, \apj, 
  687, 59
\bibitem[Khochfar \& Silk(2006)]{ks06} Khochfar, S., \& Silk, J. 2006, 
  \apj, 648, L21  
\bibitem[Kriek et al.(2006)]{kr06} Kriek, M., et al. 2006, \apj,
  649, L71
\bibitem[Kriek et al.(2007)]{kr07} Kriek, M., et al. 2007, \apj, 669,
  776
\bibitem[Kriek et al.(2008a)]{kr08a} Kriek, M., et al. 2008a, \apj,
  677, 219
\bibitem[Kriek et al.(2008b)]{kr08b} Kriek, M., van der Wel, A., van
  Dokkum, P.G., Franx, M., \& Illingworth, G.D. 2008b, \apj, 682, 896
\bibitem[Kriek et al.(2009)]{kr09} Kriek , M., van Dokkum, P. G.,
  Labb\'e, I., Franx, M., Illingworth, G. D., Marchesini, D., \& 
  Quadri, R. F. 2009, \apj, 700, 221
\bibitem[Krist(1995)]{kr95} Krist, J. 1995, in ASP Conf. Ser. 77, 
  Astronomical Data Analysis Software and Systems IV, ed. R. A. Shaw, 
  H. E. Payne, \& J. J. E. Hayes (San Francisco: ASP), 349
\bibitem[Longhetti et al.(2007)]{lo07} Longhetti, M., et al. 2007,
  \mnras, 274, 614
\bibitem[Muzzin et al.(2009)]{mu09} Muzzin, A., Marchesini, D., 
  van Dokkum, P. G., Labb\'e, I., Kriek, M., \& Franx, M. 2009, \apj, 701, 1839
\bibitem[Naab et al.(2007)]{na07} Naab, T., Johansson, P. H.,
  Ostriker, J. P., \& Efstathiou, G. 2007, \apj, 658, 710
\bibitem[Naab et al.(2009)]{na09} Naab, T., Johansson, P. H., \& Ostriker, 
  J. P. 2009, \apj, 699, L178
\bibitem[Peng et al.(2002)]{pe02} Peng, C. Y., Ho, L. C., Impey, C. D., 
  \& Rix, H.-W. 2002, \aj, 124, 266
\bibitem[Quadri et al.(2007)]{qu07} Quadri, R., et al. 2007, \aj, 134,
  1103
\bibitem[Salpeter(1955)]{sa55} Salpeter, E.E. 1955, \apj, 121, 161
\bibitem[S\'ersic(1968)]{se68} S\'ersic, J. L. 1986, Atlas de Galaxias 
  Australes (Cordoba: Obs. Astron.)
\bibitem[Taylor et al.(2009)]{ent09} Taylor, E. N. et al. 2009, ApJSS, 
  in press  (arXiv:0903.3051)
\bibitem[Toft et al.(2007)]{to07} Toft, S., et al. 2007, \apj, 671,
  285
\bibitem[Trujillo et al. (2006)]{tr06} Trujillo, I., et al. 2006,
  \apj, 650, 18
\bibitem[van Dokkum et al.(2008)]{vd08} van Dokkum, P. G. et
  al. 2008, \apj, 677, L5
\bibitem[van Dokkum et al.(2009a)]{vd09a} van Dokkum, P. G., et al. 
  2009a, PASP, 121, 2
\bibitem[van Dokkum et al.(2009b)]{vd09b} van Dokkum, P. G., Kriek, M.,
  \& Franx, M. 2009b, Nature, 460, 717
\bibitem[Williams et al.(2009)]{wi09} Williams, R. J., Quadri, R. F.,
  Franx, M., van Dokkum, P. G., Toft, S., Kriek, M., \& Labb\'e, I. 
  2009, \apj, submitted (arXiv:0906.4786)
\bibitem[Zirm et al.(2007)]{zi07} Zirm, A. W., et al. 2007, \apj, 656,
  66
\end{thebibliography}
\end{document}